%% file: main.tex
\documentclass[aps,pra,amsmath,amssymb,amsthm,floatfix,nofootinbib,notitlepage,superscriptaddress,10pt,longbibliography]{revtex4-1}
\usepackage{esvect,url, amsthm}
  
\usepackage{multirow}
\usepackage{makecell}
\pdfoutput=1

\input{definitions.tex}

\begin{document}
\title{Diffeomorphism invariant classical-quantum path integrals for Nordstrom gravity}
\author{Jonathan Oppenheim}
\affiliation{Department of Physics and Astronomy, University College London, Gower Street, London WC1E 6BT, United Kingdom}
\author{Andrea Russo}
\affiliation{Department of Physics and Astronomy, University College London, Gower Street, London WC1E 6BT, United Kingdom}
\author{Zachary Weller-Davies}
\affiliation{Department of Physics and Astronomy, University College London, Gower Street, London WC1E 6BT, United Kingdom}




\begin{abstract}
When classical degrees of freedom and quantum degrees of freedom are consistently coupled, the former diffuse, while the latter undergo decoherence. 
Here we construct a theory of quantum matter fields and Nordstrom gravity in which the space-time metric is treated classically. The dynamics is constructed via the classical-quantum path integral and is completely positive, trace preserving (CPTP), and respects the classical-quantum split. The weak field limit of the model matches the Newtonian limit of the full covariant path integral but it is easier to show that the theory is both diffeomorphism invariant, CPTP, and has the appropriate classical limit.
\end{abstract}
\maketitle

\section{Introduction}\label{sec: introduction}

The setting where gravity is effectively classical but matter retains its quantum properties belongs to the class of hybrid classical-quantum systems. These are systems whereby some parts of the system require a classical treatment while others are to be treated as quantum. Many challenges in modern physics, such as black hole evaporation, inflationary cosmology, and the measurement problem, occur in this regime, even if not all of them involve gravity directly. A good description of the classical-quantum regime is therefore paramount. While control of a quantum system through a classical variable is well understood, having a fully informative description of the classical-quantum regime hinges on properly treating the back-reaction generated by the quantum system on the classical one. In gravity, a common approach to studying back-reaction is via the semiclassical Einstein equations \cite{sato1950attempt,moller1962energy,rosenfeld1963quantization}. However, these are inconsistent when applied to all quantum states \cite{gisin1989stochastic,gisin1990weinberg,polchinski1991weinberg}, in part because they ignore the correlations between the classical and the quantum degrees of freedom~\cite{UCLpost_quantum}. It has been shown in~\cite{blanchard1993interaction,blanchard1995event,diosi1995quantum,diosi2014hybrid,kafri2014classical,tilloy2016sourcing,tilloy2017principle,poulinKITP2} that a classical-quantum treatment is possible if one allows for a stochastic coupling between the classical and quantum degrees of freedom. This stochasticity allows one to write down consistent evolution equations that are completely positive and norm preserving on the hybrid state, side-stepping the various no-go theorems and arguments about deterministic classical-quantum coupling~\cite{bohr1933on,cecile2011role,Feynman:1996kb-note,dewitt1962definition,dewitt1967quantum,eppley1977necessity,AharonovParadoxes-note,terno2006inconsistency,marletto2017gravitationally,galley2023any}. The most general form of conistent classical-quantum coupling was obtained in~\cite{UCLpost_quantum,UCLpawula}.

We call the resulting framework a ``CQ-theory" (Classical-Quantum). In this framework, the classical degrees of freedom act as a weak measurement apparatus, decohering the quantum system. The rate of this decoherence depends on the free parameters of the theory and is tied to the strength of the backreaction. The backreaction of the quantum degrees of freedom on the classical ones necessarily induces a stochastic diffusion process in the  classical phase space trajectories, transforming classically deterministic evolution into probabilistic dynamics~\cite{UCLdec_Vs_diff}.
CQ-dynamics can be expressed in multiple ways; it was originally formulated in the master equation formalism~\cite{blanchard1993interaction,diosi1995quantum,UCLpawula} but it has also been expressed as an unraveled set of stochastic PDEs~\cite{UCLunrav,UCLhealing} and via combined classical-quantum path integrals~\cite{UCLPILONG,UCLPISHORT}. In this paper, we will focus on the path integral formulation of the framework~\cite{UCLPILONG,UCLPISHORT}.

Several models of gravitational CQ theories have been studied. These include Newtonian theories based on CQ master-equations~\cite{diosi2011gravity}, measurement and feedback approaches~\cite{kafri2014classical,tilloy2016sourcing,tilloy2017principle}, as well as relativistic theories based off of both master equations \cite{UCLpost_quantum} and path integrals \cite{UCLPISHORT}. The weak field regime of gravitational CQ theories was recently explored in~\cite{UCLNewtonianLimit}. A template was provided for constructing consistent gravitational classical-quantum dynamics in the Newtonian limit. It was found that the weak field limit of the relativistic theories in~\cite{UCLpost_quantum,UCLPISHORT} took the form of \cite{tilloy2016sourcing}. These theories can be experimentally tested via the decoherence-diffusion trade-off~\cite{UCLdec_Vs_diff}, which is a relationship between the rate of decoherence, the strength of the back-reaction, and the amount of stochasticity in the theory, which all consistent classical-quantum theories must obey.

In a CQ theory of gravity, the challenge presented by the presence of Hamiltonian constraints is not trivial \cite{UCLconstraints}, even in the Newtonian limit. In the standard general relativistic treatment of the Newtonian limit, the dynamical gravitational degrees of freedom vanish, and Poisson's equation for the Newtonian potential is found from the Hamiltonian constraint. When the Newtonian limit is considered in gravitational CQ-dynamics, consistent coupling of the classical and quantum degrees of freedom necessarily introduces a diffusion process in the Newtonian potential. The effect is that of a stochastic Newtonian constraint, which requires a stochastic shift vector for the dynamics to be consistent \cite{UCLNewtonianLimit}. Away from the Newtonian limit, the gravitational constraints in CQ theories are tightly related to notions of complete positivity and diffeomorphism invariance. Specifically, while the dynamics detailed in \cite{UCLpost_quantum} are recognized as completely positive and norm-preserving, their diffeomorphism invariance was not demonstrated. Conversely, the full gravity theory quoted in \cite{UCLPISHORT} is manifestly diffeomorphism invariant. However, its consistency has yet to be fully verified because the constraint contribution to the path integral does not look completely positive. Additionally, the relationship between the dynamics of \cite{UCLpost_quantum} and~\cite{UCLPISHORT} remains unclear, partly due to the ambiguity surrounding the constraint algebra~\cite{UCLoppenheim2021constraints}. 

Since Nordstrom gravity is a self-consistent and diffeomorphism invariant theory of gravity that does not require the Hamiltonian constraints, it can sidestep these complications. Furthermore, because the weak field limit of Nordstrom gravity is the Newtonian theory, it provides us with a toy model that can be used to gain confidence in the weak field limit of~\cite{UCLpost_quantum,UCLPISHORT}. The present paper aims to construct a theory of Nordstrom gravity in the CQ framework through the path integral formalism. 
This gives a diffeomorphism invariant, though background dependent, theory of CQ-gravity, where the conformal factor plays the role of the stochastic gravitational potential. Nevertheless, we recover the same Newtonian limit behavior as in~\cite{UCLNewtonianLimit}, suggesting that the resolution of the constraints and the resulting dynamics was correct. The model also demonstrates no tension between diffeomorphism invariance and stochastic theories, nor any tension with the classical-quantum split. The theory provides a model in which various issues in quantum and classical-quantum gravity can be explored in a simpler form. It is also a manifestly Lorentz-invariant theory of stochastic collapse~\cite{grw85,Pearle:1988uh, PhysRevA.42.78, BassiCollapse,PhysRevD.34.470}, as are the models of~\cite{UCLPISHORT, UCLLorentz}. 

The paper's outline is as follows:

Section~\ref{sec: PIcq}: We summarise the CQ framework through the path integral formalism. The CQ path integral was introduced in~\cite{UCLPILONG, UCLPISHORT}, where it was shown it is possible to construct covariant path integrals for classical-quantum systems that give rise to completely positive evolution. A natural class of theories is then presented in which paths in CQ space are weighted according to a classical-quantum action, where classical paths diffuse away from their averaged equations of motion, whilst simultaneously enforcing decoherence on the quantum system. 

Section~\ref{sec: Nordstrom}:
We review Nordstrom's theory of gravity as a self-consistent theory of scalar gravity. The role of the dynamical field is played by a scalar conformal factor which evolves in a flat background. We compare the theory's advantages and shortcomings in relation to general relativity and we discuss why it is useful as a toy model of classical quantum gravity.

Section~\ref{sec: CQNord}:
We use the path integral approach to introduce and study a consistent diffeomorphism invariant theory of classical-quantum Nordstrom gravity. This provides a proof of principle that diffeomorphism invariant CQ dynamics can exist. In the Newtonian $c\to \infty$ limit we find that the theory gives rise to the Newtonian interaction on average. Due to the decoherence-diffusion trade-off, the dynamics necessarily involves diffusion away from the Newtonian solution, by an amount lower bounded by the decoherence rate into mass eigenstates. Though this example is to be understood as a toy model, it provides an instance where we have full control over the symmetries of the theory and gives support to the treatment of a more complete theory.

We conclude with a discussion in Section~\ref{sec: discussion}.

\section{Path integral for CQ dynamics}
\label{sec: PIcq}
This section introduces the CQ formalism, focusing on CQ path integrals. We outline how to define a classical-quantum state and how this state evolves through time according to completely positive CQ dynamics with interacting classical and quantum components. For a detailed description of the CQ dynamics, its derivation, and its properties, we point the reader to \cite{UCLpost_quantum,UCLpawula,UCLhealing,UCLPILONG,UCLPISHORT}.

In the CQ framework, classical degrees of freedom live in a classical configuration space $\mathcal{M}$ and are denoted by $z$. There are no restrictions on what the classical degrees of freedom can represent. The standard example is for them to be the position and momentum of a particle, in which case $\mathcal{M}= (\mathbb{R}^{2})$. On the other hand, quantum degrees of freedom live in a Hilbert space $\mathcal{H}$. We denote the set of positive semi-definite operators living in this Hilbert space as $S_{ }(\mathcal{H})$. 
We define the CQ state of the system to be described at any time by the map $\varrho(z,t): \mathcal{M} \to  S_{}(\mathcal{H})$ such that the state of the system should always admit a decomposition into its classical and quantum parts $\varrho(z,t)=p(z,t)\sigma(z,t)$, where $p(z,t)$ is understood to be the probability density over the classical degrees of freedom $z$, while $\sigma(z,t)$ is a normalized density matrix representing the quantum state. 
Therefore, the CQ state $\varrho(z)$ is at every instant subject to a normalisation constraint $\int_{\mathcal{M}} dz \Tr{\mathcal{H}}{\cqstate} =1$. To put it differently, we associate to each classical degree of freedom an un-normalized density operator $\cqstate(z)$ such that $\Tr{\mathcal{H}}{\cqstate} = p(z) \geq 0$ is a normalized probability distribution over the classical degrees of freedom and $\int_{\mathcal{M}} dz \cqstate(z) $ is a normalized density operator on $\mathcal{H}$.

The time evolution of the hybrid state has been expressed in three distinct, but ultimately equivalent, ways: via master equations in~\cite{UCLpost_quantum,UCLpawula}, unravellings~\cite{UCLunrav,UCLhealing} and path integrals~\cite{UCLPILONG,UCLPISHORT}. Regardless of one's choice, consistency requires the dynamics of the hybrid system to be linear in the density matrix, completely positive, and trace-preserving. These conditions are necessary to preserve the density matrix's statistical interpretation and give rise to positive probabilities when acting on half of an entangled state. The dynamics are also generally assumed to be time-local, which is a sensible assumption that must be made to entertain the possibility of the CQ theory describing a fundamentally classical gravitational field \cite{UCLpost_quantum}.

The original formulation of the CQ framework was carried out in the master equation picture. The advantage of this approach is that the complete positivity and the general consistency of the evolution are manifest. Unfortunately, the master equation is neither ideal when dealing with constrained systems \cite{UCLconstraints}, nor with numerical simulations. Moreover, it is hard to properly impose space-time symmetries directly at the master equation level. However, the path integral formulation of the framework works well for these tasks. First of all, constraints can be directly imposed on trajectories through the use of delta functionals, allowing one to consider only the paths of the system that live on the constraint surface. Secondly, there exist many efficient numerical methods capable of simulating path integrals. For example, those involving Monte Carlo methods find great applicability everywhere from finance~\cite{PANELLACAPUOZZO,MONTAGNA2002450,BOYLE1977323} to lattice gauge theories and quantum mechanics~\cite{CREUTZ1981427,CREUTZ1983201}. Lastly, the path integral formulation allows one to impose space-time symmetries and gauge symmetries through an action. This formalism makes it easier to formulate CQ dynamics in a covariant manner~\cite{UCLPISHORT} and to enforce the necessary principles when studying effective field theories~\cite{effectivetheories}.

The classical-quantum state can be expanded in terms of its components. Denoting a continuous quantum degree of freedom by $\phi$, we have
\begin{equation}\label{eq: statecomponents}
    \varrho(z,t) = \int d \phi^+ d\phi^- \,\varrho(z,\phi^+,\phi^-,t) \ket{\phi^+}\bra{\phi^- },
\end{equation}
where $\varrho(z,\phi^+,\phi^-,t)=\bra{\phi^+}\varrho(z,t)\ket{\phi^-}$. Here, we double the degrees of freedom in order to define the density matrix -- the $\phi^-$ field is the ket-field, while the $\phi^+$ field is the bra-field. A general CQ configuration space path integral takes the form described in~\cite{UCLPISHORT, UCLPILONG}
\begin{equation}\label{eq: transition}
    \varrho(z_f,\phi^+_f,\phi^-_f,t_f)  = \int \,\mathcal{D}z \mathcal{D} \phi^+ \mathcal{D} \phi^- \,\mathcal{N}e^{\mathcal{I}[z,\phi^+,\phi^-,t_i,t_f]}   \varrho(z_i,\phi^+_i,\phi^-_i, t_i),
\end{equation}
where $z=(z_1,...,z_n)$ are the classical degrees of freedom, $\mathcal{I}$ is the CQ action and it is implicitly understood that boundary conditions are to be imposed at $t_f$. The $\mathcal{N}$ normalisation factor is included in case the action does not preserve the norm of the state. A derivation of path integrals from CPTP master equations that find explicitly the normalization conditions on the action can be found in \cite{UCLPILONG}.

According to the main result of \cite{UCLPISHORT}, time-local CQ path integrals with actions of the form
 \begin{equation}
 \label{eq: positiveCQ}
 \begin{split}
  \mathcal{I}(z,\phi^+,\phi^-,t_i,t_f) &=  \mathcal{I}_{CQ}(z,\phi^+, t_i,t_f) + \mathcal{I}^*_{CQ}(z,\phi^-, t_i,t_f) \\
  &- \mathcal{I}_C(z, t_i,t_f) + \int_{t_i}^{t_f}dtd\vec{x}\,\sum_\gamma c^\gamma(z,x,t)L_\gamma(\phi^+)L^*_\gamma(\phi^-),
  \end{split}
\end{equation}
define completely positive CQ dynamics. In Equation~\eqref{eq: positiveCQ}, $\mathcal{I}_{CQ}$ determines the CQ interaction on each branch and $\mathcal{I}_C(z, t_i,t_f)$ is a purely classical Fokker-Plank like action~\cite{Kleinert, Sieberer_2016} taking real values. We assume that the path integral defined by Equation \eqref{eq: positiveCQ} is convergent, which enforces that $\mathcal{I}_C$ is positive (semi) definite, the real part of $\mathcal{I}_{CQ}$ is negative (semi) definite, and $c^\gamma\geq 0$. Any path integral with an action of the form in Equation \eqref{eq: positiveCQ}\ is completely positive \cite{UCLPISHORT} and the scalar gravity CQ theory we consider in Section~\ref{sec: CQNord} is of this type.
 
One can notice that in Equation~\eqref{eq: positiveCQ} $\phi^{+}\phi^{-}$ cross terms are contained only in the last term. This term is responsible for the violation of the purity of the quantum system. However, we can work in a regime where we can take the $c^\gamma=0$ so that pure quantum states are mapped to pure quantum states and there is no loss of quantum information. Conditioned on the classical trajectory, the quantum state evolution is deterministic, which provides a natural mechanism for wave-function collapse if the classical degrees of freedom are taken to be fundamentally classical. This regime coincides with the saturation of the decoherence-diffusion trade-off~\cite{UCLdec_Vs_diff}, which physically implies the presence of an exact inverse relationship between the amount of diffusion introduced in the classical system and the amount of decoherence introduced in the quantum system when the two are coupled.

A natural class of theories introduced in~\cite{UCLPISHORT, UCLPILONG} are those derivable from a classical-quantum proto-action $W_{CQ}[z,\phi]=\int dtd\vec{x}\big(\mathcal{L}_c[z]-\mathcal{V}_I[z,\phi]\big)$.  Here, and in the rest of the paper, we consider $z$ to be a configuration space variable. $\mathcal{L}_c$ is the Lagrangian density of the classical action $\mathcal{S}_c[z]=\int dtd\vec{x}\,\mathcal{L}_c[z]$ and $\mathcal{V}_I$ is the interaction potential density $V_I[z,\phi]=\int dtd\vec{x}\,\mathcal{V}_I[z,\phi]$, which in the case of general relativity is just the Lagrangian for the matter degrees of freedom. The action is given by
\begin{equation}
\label{eq: fieldConfigurationSpace2}
    \begin{split}
      \mathcal{I}(z,\phi^+,\phi^-,t_i,t_f) &=i\mathcal{S}_Q[z,\phi^+]-i\mathcal{S}_Q[z,\phi^-]+i\mathcal{S}_{FV}[z,\phi^+,\phi^-]-\mathcal{S}_{diff}[z,\phi^+,\phi^-] \\
      &=\int_{t_i}^{t_f} dt d\vec{x}\,\bigg[i\mathcal{L}_Q[z,\phi^+]-i\mathcal{L}_Q[z,\phi^-] -\frac{1}{2}\frac{ \delta  \Delta W_{CQ}}{ \delta z_{\alpha}} D_{0,\alpha\beta}[z(x)] \frac{\delta \Delta W_{CQ}}{ \delta z_{\beta}} \\& \quad\quad\quad\quad\quad\quad- \frac{1}{2} \frac{\delta \bar{W}_{CQ} }{ \delta z_\alpha} D_{2, \alpha\beta}^{-1}[z(x)] \frac{\delta \bar{W}_{CQ} }{ \delta z_\beta}
    \bigg].
    \end{split}
\end{equation}
Equation \eqref{eq: fieldConfigurationSpace2} takes the form of Equation \eqref{eq: positiveCQ} \cite{UCLPISHORT}, and is motivated by the study of canonical form CPTP master equations in \cite{UCLPILONG}.
We now explain the notation in this action.
Both $D_0^{\alpha\beta}$ and $D_{2,\alpha\beta}^{-1}$ are positive definite matrices. They are related by the so-called diffusion-decoherence trade-off~\cite{UCLdec_Vs_diff}
\begin{equation}
    4D_0^{\alpha\beta}\succeq D_{2,\alpha\beta}^{-1}.
\end{equation}
We impose the saturation of the trade-off through the matrix restriction $4D_0 = D_2^{-1}$. This ensures that the action takes the form in Equation~\eqref{eq: positiveCQ} with all the $c^\gamma=0$. Hence, the dynamics will be completely positive and the path integral will also preserve the purity of the quantum system, conditioned on the classical degrees of freedom~\cite{UCLhealing}.

In Equation~\eqref{eq: fieldConfigurationSpace2} we have introduced the notation for the $\pm$ averaged interaction
\begin{equation}
    \bar{W}_{CQ}[z,\phi] = \frac{1}{2}( W_{CQ}[z,\phi^+] + W_{CQ}[z,\phi^-]),
    \label{eq:average}
\end{equation}
and the difference in the interaction along the $\pm$ branches
\begin{equation}
    \Delta W_{CQ}[z,\phi] =  W_{CQ}[z,\phi^+] - W_{CQ}[z,\phi^-].
       \label{eq:difference}
\end{equation}
$\mathcal{L}_Q$ denotes instead the purely quantum evolution together with any interaction terms between the quantum and classical degrees of freedom. For example, it could be any quantum field theory Lagrangian. The coefficient of the difference in the interaction along the two branches is denoted as $D_0$. It regulates the decoherence of the system by suppressing paths that have different values along the $\phi^+$ branch and $\phi^-$ branches. By taking a covariant proto-action, the dynamics described by Equation \eqref{eq: fieldConfigurationSpace2} will be covariant~\cite{UCLPISHORT}. As such, they can be used to construct examples of relativistic spontaneous collapse models~\cite{grw85,Pearle:1988uh, PhysRevA.42.78, BassiCollapse,PhysRevD.34.470}. We shall see an explicit example of diffeomorphism invariant CQ dynamics in {Section~\ref{sec: CQNord}. Examples of Lorentz invariant Lindbladian dynamics can be found in~\cite{UCLLorentz}, in contrast to the Lorentz covariant dynamics found in \cite{alicki-reldecoherence} (see also~\cite{diosi2022there}).

As a toy example, we can take a single classical degree of freedom describing a scalar field $z=\{q(x)\}$ coupled to a quantum scalar degree of freedom with spatial dependence (i.e, a quantum field) 
\begin{align}
    W_{CQ}[q,\phi]=\int dt d\vec{x}\left( - \frac{1}{2}\partial_{\mu}q \partial^{\mu}q  - q(x)\phi(x) \right).
    \label{eq:SimpleExample}
\end{align}
Then, choosing for simplicity $D_0(q)^{\alpha\beta}=D_0\eta^{\alpha\beta}$, 
\begin{align}
i\mathcal{S}_{FV}[q,\phi]:=&
 -\frac{1}{2}\int dtd\vec{x}\left( \frac{ \delta  \Delta W_{CQ}}{ \delta z_{\alpha}} D_0^{\alpha \beta}(z) \frac{\delta \Delta W_{CQ}}{ \delta z_{\beta}}\right)\\
 =&
 -\frac{1}{2}D_0\int dtd\vec{x}\left(\phi^-(x)-\phi^+(x)\right)^2
 \label{eq:FVexample}
\end{align} 
acts like a Feynman-Vernon term~\cite{FeynmanVernon} which causes decoherence. The diagonal of the density matrix of Equation \eqref{eq: statecomponents} occurs when $\phi^+=\phi^-$, and on these components of the density matrix, this term does nothing. Conversely, the greater the difference between the bra and ket fields, the more the paths are suppressed by the term in Equation~\eqref{eq:FVexample}.

In a similar manner, $D_2$ tunes the averaged interaction term $\mathcal{S}_{diff}$, which is related to the diffusion of the classical system; paths that deviate from the Euler-Lagrange equations of motion, which are derived from the proto-action $W_{CQ}$, are suppressed. In the simple example of Equation \eqref{eq:SimpleExample}, one obtains
\begin{equation}
\begin{split}
  \mathcal{S}_{diff}[q,\phi]&=\frac{1}{2} \int dtd\vec{x} \left(\frac{\delta \bar{W}_{CQ} }{ \delta z_\alpha} D_{2, \alpha\beta}^{-1}(z) \frac{\delta \bar{W}_{CQ} }{ \delta z_\beta}  \right)\\
  &= \frac{1}{2D_2}\int dtd\vec{x}\left(- \partial_{\mu} \partial^{\mu}{q}(x)+\frac{\phi^+(x)+\phi^-(x)}{2}\right)^2,
  \label{eq:simpleforce}
  \end{split}
\end{equation}
where the force on the classical field is produced by the average of the bra and ket quantum fields. This term allows for fluctuations around the force but acts to suppress large deviations from them.

It is possible to recognise how the CQ path integral is connected to the path integral formulation of open quantum systems. If the action $\mathcal{I}$ was only constructed out of the quantum degrees of freedom $\mathcal{I}[\phi^+,\phi^-,t_i,t_f]=i\mathcal{S}[\phi^+,t_i,t_f]-i\mathcal{S}[\phi^-,t_i,t_f]+i\mathcal{S}_{FV}[\phi^\pm,t_i,t_f]$, we would recover the standard decoherence functional of open quantum systems. If the Feynman-Vernon term~\cite{FeynmanVernon} was not present ($\mathcal{S}_{FV}=0$), we would then recover standard unitary quantum mechanics. Much like open systems have a path integral formulation of their master equation version, the CQ path integral can be directly thought of as coming from the master equation formulation~\cite{UCLpost_quantum}. Nevertheless, according to~\cite{UCLPISHORT}, we can take the path integral as the starting point of the CQ framework. This allows for a simpler definition of the path integral, given that deriving a clean form from the master equation requires one to map between configuration space and canonical co-ordinates \cite{UCLPILONG}.

In this section, we have introduced the formalism of the CQ framework. We have summarised how a positive path integral version of the dynamics can be written and interpreted and how the Classical-Quantum path integral incorporates the back-reaction between the quantum and classical systems, leading to decoherence and diffusion effects. In~\cite{UCLhealing}, it was also shown that when conditioned on the classical degrees of freedom a path integral that saturated the decoherence-diffusion tradeoff~\cite{UCLdec_Vs_diff} preserves the purity of the quantum system.  In the next section, we will apply it to a diffeomorphism invariant theory of scalar classical-quantum gravity and study its weak field limit. We will find that, in the Newtonian limit, the theory predicts diffusion about the standard semiclassical Newtonian solution by an amount that is lower bounded by the decoherence rate into mass eigenstates according to~\cite{UCLNewtonianLimit}.

\section{Nordstrom gravity}\label{sec: Nordstrom}

So far, we have reviewed how to couple classical and quantum degrees of freedom via CQ path integrals. No explicit choice of classical and quantum systems has been made yet, and the equations presented so far hold in general. We will now discuss the classical system of interest in this paper, before integrating it into the CQ framework in Section~\ref{sec: CQNord}.

Nordstrom gravity \cite{Nord, ravndal2004scalar} was a first attempt at merging Newtonian gravity with relativity and ultimately led to the formulation of GR as it currently stands \cite{Norton}. In its final formulation, Nordstrom Gravity can be thought of as a self-consistent scalar theory of gravity. It was the first metric theory of gravity, hence it obeys the equivalence principle. The classical theory is described through a conformally flat spacetime background which couples to matter via the equations:

    \begin{align}
        &\mathcal{R}=\frac{24\pi G}{c^4} T \label{eq: nordstromEq}, \\
        & C_{\mu\nu\rho\sigma}=0, \label{eq: Weyl}
    \end{align}
where $\mathcal{R}$ is the Ricci scalar, $T$ denotes the trace of the stress-energy tensor for the matter degrees of freedom $\phi_m$ and $C_{\mu \nu \rho \sigma}$ is the Weyl tensor.  Equation~\eqref{eq: nordstromEq} is the dynamical equation of motion linking the Ricci scalar to the trace of the stress-energy tensor. The vanishing of the Weyl tensor in Equation~\eqref{eq: Weyl} implies that the metric is conformally flat and always takes the form:
\begin{equation}
   \label{eq: conformal}
   g_{\mu\nu}=e^{\frac{2\Phi}{c^2}}\eta_{\mu\nu},
\end{equation}
 where $\Phi$ is the conformal degree of freedom and $\eta_{\mu \nu}$ is the Minkowsky metric. 

Nordstrom gravity merges relativistic ideas of causality with Newtonian gravity but lacks many of the properties required for a full description of gravitational phenomena. One can immediately notice that the conformal metric couples only to the \emph{trace} of the stress-energy tensor. However, some forms of energy and momentum, like the stress-energy tensor for electromagnetic radiation, are traceless. Therefore, Nordstrom gravity lacks the ability to describe some gravitational phenomena. In particular, it 
does not predict the bending of light, in direct contrast with gravitational lensing effects observed by astronomers \cite{Dyson:1920cwa, Bartelmann_2010}. With regard to other effects, Nordstrom gravity correctly predicts the result of the Pound-Rebka experiment for gravitational frequency shift but fails to predict the correct time delay factor and is missing subleading corrections to the acceleration of static test particles.
 
However, much like General Relativity, Nordstrom's theory is \textit{diffeomorphism invariant}, by which we mean that $(g,\Phi,\phi_m)$ is a solution to the equations of motion if and only if $(g^*, \Phi^*, \phi_m^*)$ is also a solution to the same equation's of motion, where $^*$ denotes the transformed variables after a diffeomorphism. In particular, conformal flatness is preserved under diffeomorphisms. The conformal flatness condition does not fix the conformal factor, which is the dynamical gravitational degree of freedom. For example, given the form of the metric in Equation~\eqref{eq: conformal}, the Ricci scalar reads
    \begin{equation}
        \mathcal{R}=-\frac{6\tilde{\square}e^{\frac{\Phi}{c^2}}}{c^2}e^{-\frac{3\Phi}{c^2}},
    \end{equation}
where $\tilde{\square}=\partial_\mu\partial_\nu\eta^{\mu\nu}$ is the \emph{flat} space D'Alabertian. In the vacuum state ($T=0$) the field equation is the wave equation for the scalar field $\tilde{\square}e^{\frac{\Phi}{c^2}}=0$. Therefore, the theory has a propagating conformal scalar degree of freedom, but this kind of gravitational wave differs from those predicted by General Relativity as they are scalar waves and do not have a spin-2 mode.

Nonetheless, despite being diffeomorphism invariant, the Nordstrom theory is \textit{intuitively} background dependent. It has a preferred frame given by the Minkowski metric due to the condition that the metric be conformally flat. In particular, it admits a background-dependent formulation (which is still diffeomorphism invariant) where we first stipulate that the metric takes the form of Equation~\eqref{eq: conformal}, with the dynamics determined by Equation~\eqref{eq: nordstromEq}. For a more detailed discussion on the relationship between diffeomorphism invariance and background independence, we refer the reader to \cite{Pooley}. We  now study a CQ version of Nordstrom gravity. This provides a diffeomorphism invariant, self-consistent theory of CQ gravity that has a sensible Newtonian limit.

\section{CQ Path Integral for Nordstrom Gravity}
\label{sec: CQNord}

In this section, we will first introduce CQ gravitational path integrals for general relativity, outlining the tension between complete positivity of the dynamics and the gravitational constraints. Then, we construct the CQ path integral for Nordstrom gravity.  Our choice of classical system will be the space-time metric, while the matter degrees of freedom will have a quantum nature.
\subsection*{CQ general relativity}
When writing the path integral for General relativity, we follow~\cite{UCLPISHORT} and write a manifestly covariant path integral over 4-geometries $g$, of the form given by Equation~\eqref{eq: fieldConfigurationSpace2}
\begin{equation}
\label{eq: GRpathintegral}
   \varrho(\Sigma_f,\phi^+_f,\phi^-_f,t_f) = \int\mathcal{D}g
    \mathcal{D} \phi^+\mathcal{D} \phi^-\,\mathcal{N}  \ e^{ \mathcal{I}_{CQ}[g,\phi^+,\phi^-,t_i,t_f] }\varrho(\Sigma_i,\phi^+_i,\phi^-_i,t_i).
\end{equation}
with:
\begin{equation}
\label{eq: CQGRaction}
    \begin{split}
\mathcal{I}_{CQ}[g,\phi^\pm] &=  \int_{t_i}^{t_f} dtd\vec{x}\,\bigg[ i\big(\mathcal{L}_{\mathcal{Q}}[g,\phi^+] - \mathcal{L_{\mathcal{Q}}}[g,\phi^-] \big) \\
&\quad\quad\quad\quad -\frac{\text{Det}[-g] }{8}\big(T^{\mu\nu}[\phi^+]-T^{\mu\nu}[\phi^-] \big)D_{0,\mu\nu\rho\sigma}[g]\big(T^{\rho\sigma}[\phi^+]-T^{\rho\sigma}[\phi^-] \big)\\
&\quad\quad\quad\quad- \frac{\text{Det}[-g]|c^8}{128\pi^2 G^2_N}\left(G^{\mu\nu}-\frac{8\pi G_N}{c^4}\bar{T}^{\mu\nu}[\phi^+,\phi^-]\right)D_{0,\mu\nu\rho\sigma}[g]\left(G^{\rho\sigma}-\frac{8\pi G_N}{c^4}\bar{T}^{\rho\sigma}[\phi^+,\phi^-]\right)\bigg]
  \end{split}
\end{equation}
where $\mathcal{L}_{Q}[g,\phi^\pm]$ is the Lagrangian for the quantum matter in curved spacetime and we have suppressed the metric dependence of all terms in Einstein equations for clarity. $\Sigma_i$ and $\Sigma_f$ are the initial and final spatial surfaces. The term $\bar{T}^{\mu\nu}[\phi^+,\phi^-]$ indicated the average of the bra and ket stress-energy tensors, as in Equation~\eqref{eq:difference}
\begin{equation}
 \bar{T}^{\mu\nu}[\phi^+,\phi^-]=\frac{1}{2}(T^{\mu\nu}[\phi^+]+T^{\mu\nu}[\phi^-]).   
\end{equation}
Here we have assumed that the decoherence-diffusion trade-off is saturated $(4D_0=D_2^{-1})$. This is a manifestly diffeomorphism invariant hybrid path integral for General Relativity, and it is fully characterised by the tensor density $D_{0,\mu\nu\,\rho\sigma}[g]$. As explained in~\cite{UCLPISHORT}, if we choose $D_0$ such that it is positive semi-definite, we would have a consistent treatment of semiclassical General Relativity. Unfortunately, choosing a positive semi-definite $D_0$, and capturing the transverse part of the Einstein equations is not possible in Lorentzian signature. For example, one can consider 
    \begin{equation}
        \label{eq: transversediffusion}
        D_{0,\mu\nu\rho\sigma}=\frac{D_0(x)}{2\sqrt{-g}}(g_{\mu\rho}g_{\nu\sigma}+g_{\nu\rho}g_{\mu\sigma}-2\beta g_{\mu\nu}g_{\rho\sigma}),
    \end{equation}
with $D_0$ a positive constant. In Lorentzian signature, this is not positive semi-definite. 
However, one can choose
    \begin{equation}
        \label{eq: tracediffusion}
        D_{0,\mu\nu\rho\sigma}=\frac{D_0(x)}{\sqrt{-g}}g_{\mu\nu}g_{\rho\sigma}.
    \end{equation}
This will lead to a positive semi-definite path integral describing suppressed trajectories as they diffuse away from the \textit{trace} of Einstein's equations. 

Alternatively, if one is happy with a consistent diffeomorphism invariant toy model of gravity in the CQ framework, one can instead start with a scalar theory of gravity. In particular, Nordstrom gravity is an ideal candidate as a self-consistent theory of gravity that allows us to study the gravitational backreaction of spacetime and quantum matter, without worrying about the constraints of General relativity or the positivity of the full CQ path integral for GR.

\subsection*{CQ Nordstrom}
To construct the CQ theory of Nordstrom gravity, we let $\phi_m$ denote the quantum matter degrees of freedom such that $\mathcal{L}_Q[g,\phi_m]$ is the matter Lagrangian (inclusive of the appropriate metric determinant factor). Nordstrom gravity can be derived classically from the action principle in the Jordan frame defined in~\cite{Nordstromaction}, which we summarise in Appendix~\ref{app: Nordconstrclass}. In that derivation, a Lagrange multiplier is used to impose the conformal flatness of the spacetime~\eqref{eq: conformal}. Here, we are faced with two choices. We could insert the constraint in the proto-action directly or impose it through a delta functional. Given that in Nordstrom gravity matter fields do not couple to the Weyl tensor, we do not expect the backreaction of the quantum degrees of freedom to break the conformal flatness of the metric. After some consideration, one can realise that it is more sensible to choose the latter, imposing the constraint in a way that is more akin to a gauge fixing of the classical degrees of freedom. Therefore, we construct the proto-action for Nordstrom gravity with matter as
    \begin{equation}
    \label{eq: Nordstrom_action}
        W_{CQ}[g_{\mu\nu},\phi_m]=-\frac{c^4}{48\pi G_N}\int d^4x\sqrt{-g}\,\mathcal{R}+\int d^4x\,\mathcal{L}_Q[g,\phi_m].
    \end{equation}
While this action might look similar to the Einstein-Hilbert action, one should notice the different coefficients of the gravitational sector and the different relative signs between the gravitational and matter part, both are required to obtain the correct Nordstrom dynamical equation.
We are now in the position to write down the CQ path integral. We choose the trace realisation of the decoherence coefficient in Equation~\eqref{eq: tracediffusion} and impose the conformal flatness constraint as a delta functional through a Lagrange multiplier $\lambda_\mu^{\nu\rho\sigma}$
\begin{equation}
\label{eq: NordPathsetup}
\begin{split}
  \varrho(\Sigma_f,\phi^+_{m,f},\phi^-_{m,f},t_f) & = \int \mathcal{D}g\mathcal{D}\phi_m^+\mathcal{D}\phi_m^- \mathcal{D}\lambda_\mu^{\nu\rho\sigma}\,\mathcal{N}\,e^{\mathcal{I}_{CQ}(g_{\mu\nu},\phi^+_{m},\phi^-_{m},t_f,t_i)} \varrho(\Sigma_i,\phi^+_{m,i},\phi^-_{m,i},t_i),
  \end{split}
\end{equation}
where the CQ action is
    \begin{equation}
    \begin{split}
        \mathcal{I}_{CQ}(g_{\mu\nu},\phi^+_{m},\phi^-_{m},t_f,t_i)=\int_{t_i}^{t_f} dtd\vec{x}\bigg[ & \;i(\mathcal{L}_Q[g, \phi_m^+]- \mathcal{L}_Q[g,\phi_m^-])-\frac{D_0(x)}{2\sqrt{-g}}\frac{\delta\Delta W_{CQ}}{\delta g^{\mu\nu}} g_{\mu\nu}g_{\rho\sigma}\frac{\delta\Delta W_{CQ}}{\delta g^{\rho\sigma}}\\
        &- \frac{2D_0(x)}{\sqrt{-g}}\frac{\delta\bar{W}_{CQ}}{\delta g^{\mu\nu}} g_{\mu\nu}g_{\rho\sigma}\frac{\delta\bar{W}_{CQ}}{\delta g^{\rho\sigma}} -i\lambda_\mu^{\nu\rho\sigma}C^\mu_{\nu\rho\sigma}\bigg],
    \end{split}
    \end{equation}
and we have saturated the decoherence diffusion tradeoff:
    \begin{equation}
        \label{eq: decodiff}
        4D_0[g]=D_2^{-1}[g].
    \end{equation}

Once we integrate over the Lagrange multiplier, the delta function will ensure that we only sum over 4-geometries that are conformally flat $C_{\nu \rho \sigma}^{\mu}=0$. Any conformally flat metric can be written as  $g_{\mu\nu}=e^{2\frac{\Phi}{c^2}}\eta_{\mu\nu}$ for some $\Phi$ by definition. Therefore, we find the Nordstrom hybrid path integral

\begin{equation}\label{eq: NordPathFull}
\begin{split}
  \varrho(\Sigma_f,\phi^+_{m,f},\phi^-_{m,f},t_f)  = \int \mathcal{D}\Phi \mathcal{D}\phi_m^+\mathcal{D}\phi_m^-\,\mathcal{N}\, \exp &\bigg[   \int dtd\vec{x} \;i(\mathcal{L}_Q[\phi_m^+]- \mathcal{L}_Q[\phi_m^-]) -\frac{\sqrt{-g} D_0(x)}{8}\big( T[ \phi_m^+] - T[\phi_m^-]\big)^2 \\
  &\quad -\frac{\sqrt{-g} c^8 D_0(x)}{1152\,\pi^2 G_N^2 }\left(\mathcal{R}-\frac{24 \pi G_N}{c^4} \bar{T}[ \phi_m^+,\phi_m^-]\right)^2 \bigg]\varrho(\Sigma_i,\phi^+_{m,i},\phi^-_{m,i},t_i),
  \end{split}
\end{equation}
where we have suppressed the $\Phi$ dependence in $\mathcal{R}$, $T$ and $\mathcal{L}_Q$ to lighten the notation. When integrating over conformally flat metrics, we include any Jacobian factor in the measure $ \mathcal{D}g^{C=0} \sim \frac{2}{c^2} e^{\frac{2 \Phi}{c^2}} \mathcal{D}\Phi$, as it will not be relevant to the Newtonian limit of interest in this paper. In particular, to leading order, we have that $Dg^{C=0} \sim \frac{2}{c^2} \mathcal{D}\Phi$. Since the action in Equation \eqref{eq: NordPathFull} contains quantum terms proportional to the square of the stress-energy tensor, a sufficient condition for the path integral to be normalized is that the purely quantum part of the action $\mathcal{L}_{\mathcal{Q}}[q,\phi^\pm]$ contains higher derivative kinetic terms $\sim \ddot{\phi}^2$ \cite{ZWDquantum2024}, which is suggestive that Equation \eqref{eq: NordPathFull} describes an effective theory \cite{effectivetheories}.

The action has the effect of diffusing away from the bra/ket averaged Nordstrom equations, whilst simultaneously decohering the quantum system according to the stress-energy tensor of the matter and the coupling $D_0[\Phi]$. Treated classically, the action is manifestly diffeomorphism invariant, which also includes the case where the diffeomorphism $\sigma: M \to M $ is dependent on the classical and quantum trajectories $\sigma[\Phi, \phi^{\pm}_m]$. However, just as for the classical theory, the CQ theory is not background-independent, and has a preferred frame given by the requirement that the metric is conformally flat. 

We now wish to compute the Newtonian limit of the theory in order to gain insight into the Newtonian limit of more general classical-quantum theories and compare and contrast it with~\cite{UCLNewtonianLimit}. We recall that the final goal is to search for low-energy experimental signatures of CQ by treating the gravitational field classically. To that end, we shall take the quantum degrees of freedom to be described by a pressureless dust distribution $\hat{T}^{\mu\nu}=\hat{m}(x)U^\mu U^\nu$ where $\hat{m}(x)^{\pm}$ is a (smeared) quantum mass density. 

Using the conformally flat metric and our choice of matter we can rewrite some of the quantities in the path integral as:
    \begin{equation}
        \sqrt{-g}=c\,e^{\frac{4\Phi}{c^2}} , \ \mathcal{R}[\Phi]=-\frac{6\tilde{\square}e^{\frac{\Phi}{c^2}}}{c^2}e^{-\frac{3\Phi}{c^2}}, \ T^{\pm}[\Phi,m^{\pm}]=-e^{\frac{2\Phi}{c^2}} m^{\pm}(x) ,
    \end{equation}
and the path integral takes the form 
\begin{equation}
\label{eq: NordPathFinal}
\begin{split}
  \varrho(\Sigma_f,m^+_f,m^-_f,t_f)  = \int \mathcal{D}\Phi \mathcal{D}m^+\mathcal{D}m^- \mathcal{N}\exp&\left[\int_{t_i}^{t_f} dtd\vec{x} \;i\big(\mathcal{L}_Q[\Phi, m^+]- \mathcal{L}_Q[\Phi,m^-]\big) -\frac{cD_0(x)e^{\frac{6\Phi}{c^2}}}{8}\big( m^+(x)-m^-(x)\big)^2\right. \\
  &\quad\left. - \frac{c^7 D_0(x)}{192\,\pi^2 G_N^2 }\left(-e^{\frac{\Phi}{c^2}}\tilde{\square}e^{\frac{\Phi}{c^2}}+\frac{4 \pi G_Ne^{\frac{6\Phi}{c^2}}}{c^2}\bar{m}(x)\right)^2 \right]\varrho(\Sigma_i,m^+_i,m^-_i,t_i),
  \end{split}
\end{equation}
where $\bar{m}(x)=\frac{1}{2}\big( m^+(x)+m^-(x)\big)$.

We take the Newtonian $c\to \infty$ limit of the metric perturbations. Carrying out the transformations, we get
    \begin{align}
        &\sqrt{-g}=c\; e^{\frac{4\Phi}{c^2}} \approx c\left(1+ \frac{4 \Phi}{c^2}\right) + \mathcal{O}\left(\frac{1}{c^3}\right),\\
        &e^{\frac{\Phi}{c^2}}\tilde{\square}e^{\frac{\Phi}{c^2}}\approx\frac{\tilde{\square}\Phi}{c^2}+\mathcal{O}\left(\frac{1}{c^4}\right), \\
        &\frac{4 \pi G_N e^{\frac{6\Phi}{c^2}}}{c^2}\bar{m}(x) \approx \frac{4 \pi G_N}{c^2 }\bar{m}(x) + \mathcal{O}\left(\frac{1}{c^4}\right).
    \end{align}

To leading order in $c$, we then arrive at the Newtonian limit of the CQ scalar gravity theory 
\begin{equation}
    \label{eq: PathNewt}
        \varrho(\Sigma_f,m^+_f,m^-_f,t_f)= \int \mathcal{D} \Phi\mathcal{D}m^+\mathcal{D}m^-  \,\mathcal{N}\,e^{\mathcal{I}_{CQ}[\Phi,m^+,m^-,t_i,t_f]}  \varrho(\Sigma_i,m^+_i,m^-_i,t_i), 
    \end{equation}
with CQ action given by:
\begin{equation}
\label{eq: NordPathNewt}
\begin{split}
    \mathcal{I}_{CQ}[\Phi,m^+,m^-,t_i,t_f] = \int_{t_i}^{t_f} dtd\vec{x}\, &\bigg[i\big(\mathcal{L}_{Q}[m^+]-\mathcal{V}_I[\Phi,m^+]-\mathcal{L}_{Q}[m^-]+\mathcal{V}_I[\Phi,m^-]\big)\\& -\tilde{D}_0(x)\big( m(x)^+ - m(x)^-\big)^2 - \frac{c^2\tilde{D}_2^{-1}(x)}{6} \left( \frac{\tilde{\square} \Phi}{4 \pi G } - \bar{m}(x)\right)^2 \bigg].
\end{split}
\end{equation}
where $\mathcal{V}_I[\Phi,m^\pm]$ is the interaction potential coming from the expansion of the metric determinant in the quantum Lagrangian's and we have defined $ \tilde{D}_0(x) = \frac{cD_0(x)}{8}$, $\tilde{D}_2^{-1}(x) =\frac{cD_0(x)}{2} $ which are related to the decoherence and diffusion coefficients of the Newtonian potential.
With these re-definitions, these coefficients relate to physically observable quantities: $\tilde{D}_0$ quantifies the suppression of quantum trajectories away from $m(x)^+ = m(x)^-$, which is the decohered trajectory. On the hand hand $\tilde{D}_2(x) = \frac{1}{2}\sigma^2_{\Phi}$ where $\sigma_{\Phi}$ quantifies the variance away from the semiclassical Newtonian solution. 
They saturate the decoherence-diffusion relation:
\begin{equation}
\label{eq: tradeoffnordstrom}    
4 \tilde{D}_0(x) = \tilde{D}_2^{-1}(x).
\end{equation}
In Equation~\eqref{eq: NordPathNewt}, we have explicitly kept the d'Alambert operator to highlight the fact that, differently from~\cite{UCLNewtonianLimit}, $\phi$ is in principle still a dynamical variable as it is not constrained by the ADM constraints. Keeping the d'Alambert operator is also required for normalisation \cite{ZWDquantum2024}. However, in the slow-moving limit, we recover the randomly sourced Poisson equation and exactly match the Newtonian limit of~\cite{UCLNewtonianLimit}.

Although the scalar theory is a toy model, it is worth highlighting some of its appealing features which we expect to apply to more general CQ theories. Firstly, as mentioned, we have both diffusion in the Newtonian potential and decoherence in the quantum system, by an amount quantified by Equation~\eqref{eq: tradeoffnordstrom}. More generally we expect that the amount of diffusion in the Newtonian potential will be lower bounded by~\eqref{eq: tradeoffnordstrom} which is an experimental signature of classical-quantum theories~\cite{UCLdec_Vs_diff}. Indeed, we see the same decoherence-diffusion relation between the diffusion of the Newtonian potential away from its averaged solution, and the decoherence rate into the mass eigenbasis, as for the Newtonian path integral of~\cite{UCLNewtonianLimit}. 

Secondly, expanding out Equation's~\eqref{eq: NordPathFull} and~\eqref{eq: NordPathNewt}, we see that all the ${\pm}$ cross terms cancel so that the path integral preserves the purity of the quantum state even though the state decoheres into the mass eigenbasis. Such classical quantum theories, therefore, provide a natural mechanism to describe wavefunction collapse via the interaction of a classical field with a quantum one. Moreover, the fact that the classical field is dynamical is enough to restore apparent diffeomorphism invariance in the theory, which can be seen via the diffeomorphism invariant action in~\eqref{eq: NordPathFull}. The fact that gravity interacts with matter through its stress-energy tensor provides an amplification mechanism by which small masses can maintain coherence whilst macroscopic objects will be decohered. Indeed, if one disregards the classical degrees of freedom, then the resulting dynamics are very similar to the dynamics of collapse theories~\cite{grw85,Pearle:1988uh, PhysRevA.42.78, BassiCollapse,PhysRevD.34.470} but we see that the full theory is diffeomorphism invariant due to the fact that we consider a dynamical classical field.

Let us finally comment on the continuity properties of the Nordstrom theory and its Newtonian limit. Typically in path integral approaches the path integral can be understood as an integral over paths that are (almost surely) continuous. The reason for this is that they typically involve kinetic terms $\frac{i}{\hbar}[\frac{1}{\delta t}(x_{t+\delta t} - x_t)]^2$ which give a highly oscillatory contribution to the path integral $\sim e^{\frac{i}{\hbar \delta t^2}}$ if the paths are discontinuous. For the Nordstrom theory, we expect similar behaviour for the gravitational field in the full path integral (Equation \eqref{eq: NordPathFull}) due to the $-\frac{\mathcal{R}^2}{4 D_2}$ term which includes kinetic terms through $\tilde{\square}\Phi$. However, in the $c\to \infty$ limit, we are led to neglect such terms, leading to a discontinuous path integral in Equation \eqref{eq: NordPathNewt}. This is a remnant of the approximation, and we expect that any physical measurable quantity of interest should be smeared over a time scale to reflect this.

\section{Discussion}
\label{sec: discussion}
In this work, we have constructed the covariant path integral of a full diffeomorphism invariant theory of CQ Nordstrom gravity and derived its Newtonian limit. The result matches the Newtonian limit obtained in~\cite{UCLNewtonianLimit}, where we started from general relativity and proceeded with gauge fixing the Newtonian metric. In both cases, the non-relativistic path integral describes a classical Newtonian gravitational field diffusing around Poisson's equation of motion. At the same time, quantum matter degrees of freedom decohere into mass eigenstates due to the backreaction of the classical geometry. In order for the dynamics to be completely positive, the amount of diffusion is lower bounded by the coherence time for superpositions of mass distributions. 
Differently from~\cite{UCLNewtonianLimit}, our choice of classical system was Nordstrom's scalar theory of gravity. While Nordstrom's theory does not accurately describe all gravitational phenomena, it is nonetheless a self-consistent theory of relativistic gravitation. The main appeal of this choice is that it allows us to bypass any discussion or concern regarding gravitational constraints, serving as proof that there is no fundamental impediment to constructing a positive diffeomorphism invariant theory of CQ gravity. We expect the scalar theory to be an interesting toy model for exploring conceptual issues around classical-quantum gravity theories. For example, whether or not they can be made renormalizable \cite{UCLrenorm}. 

One can also study the Nordstrom theory as a toy model for exploring experimental signatures of CQ gravity since the theory has the same Newtonian limit \cite{UCLNewtonianLimit}. Tests of CQ gravity include proposals for the detection of gravitationally induced entanglement between masses in interferometric setups~\cite{kafri2013noise,kafri2015bounds,bose2017spin,MarlettoVedral2017,Marshman_2020,pedernales2021enhancing,christodoulou2022locally,Danielson:2021egj},~\cite{carney2021testing}, c.f.~\cite{streltsov2022significance},which may become feasible in the next decade or two. 
There are also experiments that can exclude classical theories of spacetime through the  decoherence vs diffusion trade-off~\cite{UCLdec_Vs_diff}, as in Equation~\eqref{eq: tradeoffnordstrom}.
So far, the upper bounds on diffusion have been found from precision mass experiments measuring the variation in acceleration experienced by mass in a Newtonian potential. Tighter bounds would follow from increased precision in the readings of acceleration experienced by the masses. 
On the other hand, improved decoherence times would lead to stronger lower bounds on $D_2$ and further squeeze theories in which gravity remains classical. As explained in detail in~\cite{UCLdec_Vs_diff}, since the trade-off is in terms of the inverse Lindbladian coupling $D_0^{-1}$, one can also constrain classical theories of gravity by bounds on anomalous heating of the quantum system~\cite{bps,ghirardi1986unified,ballentine1991failure,pearle1999csl,bassi2005energy,adler2007lower,lochan2012constraining,nimmrichter2014optomechanical,bahrami2014testing,laloe2014heating,bahrami2014proposal,goldwater2016testing,tilloy2019neutron,donadi2020underground}. 
Other experimental proposals, look for coherence or correlations in the gravitational interaction~\cite{lami2023testing,kryhin2023distinguishable}. The variety of experimental proposals as well as new theoretical tools, suggest that probing the quantum vs classical nature of spacetime can be accomplished at low energy and is likely to shed light on attempts to reconcile quantum theory with general relativity.

\section*{Acknowledgements}
We would like to thank Isaac Layton, Emanuele Panella, Maite Arcos and Andrew Svesko for valuable discussions. We especially thank Bill Unruh who suggested we look at Nordstrom gravity as an interesting toy model. 
JO is supported by an EPSRC Established Career Fellowship, and a Royal Society Wolfson Merit Award, A.R acknowledges financial support from EPSRC. This research was supported by the National Science Foundation under Grant No. NSF PHY11-25915 and by the Simons Foundation {\it It from Qubit} Network.


\appendix

\section{Nordstrom gravity from action variation}\label{app: Nordconstrclass}
In this appendix, we summarise the derivation of the Nordstrom equation of motion from an action principle. This derivation is discussed in detail in \cite{Nordstromaction}. To begin, one considers an action similar to that of Equation~\eqref{eq: Nordstrom_action}
\begin{equation}
    \mathcal{S}[g_{\mu\nu},\phi_m,\lambda_\mu^{\nu\rho\sigma}]=-\frac{c^4}{48\pi G_N}\int d^4x\sqrt{-g}\left(\mathcal{R}+\lambda_\mu^{\nu\rho\sigma}C^\mu_{\nu\rho\sigma}\right)+\int d^4x \, \mathcal{L}_Q[g,\phi_m],
\end{equation}
where $\lambda_\mu^{\nu\rho\sigma}C^\mu_{\nu\rho\sigma}$ is the Weyl tensor which is constrained to vanish through the Lagrange multiplier $\lambda_\mu^{\nu\rho\sigma}$. Here the Lagrange multiplier is directly inserted in the action. When this action is varied with respect to the matter degrees of freedom, the Lagrange multiplier and the metric, one obtains for its extrema:
    \begin{equation}
    \label{eq: Lagr_variation}
        \frac{\delta \mathcal{S}}{\delta \phi_m}=\nabla_\nu T_m^{\mu\nu}=0, 
 \quad \frac{\delta \mathcal{S}}{\delta \lambda_\mu^{\nu\rho\sigma}}=C^\mu_{\nu\rho\sigma}=0 ,
    \end{equation}
which impose energy conservation and conformal flatness of the metric on the constraint surface $g_{\mu\nu}-e^{\frac{2\Phi}{c^2}}\eta_{\mu\nu}\approx 0$. Lastly,
    \begin{equation}
        \frac{\delta \mathcal{S}}{\delta g^{\mu\nu}}=\begin{cases} \mathcal{R}-\frac{24\pi G_N}{c^4}T_m=0 & \text{Trace part} \\ \frac{\partial^\alpha\partial_\beta\lambda_{\mu\alpha\nu}^\beta}{\phi^2}=-\frac{24\pi G_N}{c^4}\left(T_{m,\mu\nu}-\mathcal{R}_{\mu\nu}-\frac{1}{4}g_{\mu\nu}(T_m+\mathcal{R})\right) & \text{Traceless part}\end{cases}
    \end{equation}
 This leaves the equation of Jordan's frame version of Nordstrom's final theory to be
\begin{equation}
    \mathcal{R}-\frac{24\pi G_N}{c^4}T_m=0, \quad \nabla_\nu T_m^{\mu\nu}=0, 
 \quad C^\mu_{\nu\rho\sigma}=0.
\end{equation}
The traceless part can be written in terms of flat covariant derivatives using the conformal flatness of the metric
\begin{align}
    \label{eq: Riccinordstrom}
        &\mathcal{R}_{\mu\nu}=-e^{-\frac{\Phi}{c^2}}\partial_\mu\partial_\nu e^{\frac{\Phi}{c^2}}-\eta_{\mu\nu}\,e^{-\frac{\Phi}{c^2}}\tilde{\square}e^{\frac{\Phi}{c^2}}+4e^{-\frac{2\Phi}{c^2}}\partial_\mu e^{\frac{\Phi}{c^2}}\partial_\nu e^{\frac{\Phi}{c^2}}-\eta_{\mu\nu}\,e^{-\frac{2\Phi}{c^2}} \partial_\rho\partial^\rho e^{\frac{\Phi}{c^2}}\\
        &\mathcal{R}=-\frac{6\tilde{\square}e^{\frac{\Phi}{c^2}}}{c^2}e^{-\frac{3\Phi}{c^2}}.
    \end{align}
This is often found in the literature written using the notation $e^{\frac{\Phi}{c^2}}=\phi$, which results in
\begin{align}
        &\mathcal{R}_{\mu\nu}=-\frac{\partial_\mu\partial_\nu\phi}{\phi}-\eta_{\mu\nu}\frac{\tilde{\square}\phi}{\phi}+4\frac{\partial_\mu\phi\partial_\nu\phi}{\phi^2}-\eta_{\mu\nu}\frac{\partial_\rho\partial^\rho\phi}{\phi^2},\\
        &\mathcal{R}=-\frac{6\tilde{\square}\phi}{c^2\phi^3}.
\end{align}
Once the solution for the conformal factor $\phi$ is known, the traceless equation is an equation of motion for the Lagrange multiplier $\lambda_{\mu\nu\rho\sigma}$. However, the system is undefined as $\lambda$ has 10 components (it has the same symmetries as Weyl tensor), but the system of equation composed of~\eqref{eq: Lagr_variation} and~\eqref{eq: Riccinordstrom}, being traceless, has only 9. Classically, this is not a problem as the Lagrange multiplier does not enter the equation of motion for the Nordstrom field $\phi$.
For the CQ equations considered in Section \ref{sec: CQNord}, the Lagrange multiplier enforces a delta function $\delta C$ in the path integral. This allows us to only consider diffusion away from the trace of the equations of motion, which are exactly the Nordstrom field equations.

\bibliography{NewtLimitbib.bib}

\end{document}

%% file: definitions.tex
\usepackage{amsthm}
\usepackage{latexsym}
\usepackage{amsfonts}
\usepackage{bbm,dsfont}
\usepackage{graphicx,xcolor}
\usepackage{braket}
\usepackage{hyperref}
\usepackage[normalem]{ulem}
\usepackage{braket}
\definecolor{bottle_green}{RGB}{0,106,78}
\definecolor{celadon_green}{RGB}{47,132,124}
\definecolor{emerald}{RGB}{80,220,100}
\definecolor{jade}{RGB}{0,168,107}
\hypersetup{colorlinks,
linkcolor=bottle_green,
citecolor=celadon_green,
urlcolor=emerald,
anchorcolor=jade}

\newcommand{\Tr}[2]{\mathrm{Tr}_{#1}\left[ {#2} \right]}

\newcommand{\cqstate}{\varrho}

\newcommand{\beq}{\begin{equation}}
\newcommand{\eeq}{\end{equation}}